\documentclass[aps,prl,twocolumn,superscriptaddress]{revtex4-1}
\usepackage{natbib}
\usepackage{amsmath}
\usepackage{graphicx}

\usepackage{bm}%
\usepackage[colorlinks=true,linkcolor=blue]{hyperref}%

\usepackage{ulem}
\begin{document}

\title{Grain boundary melting in ice}

\author{E. S. Thomson}
\email[corresponding author: ]{erik.thomson@chem.gu.se}
\affiliation{Department of Chemistry \& Molecular Biology, Atmospheric Science, University of Gothenburg, 41296 Gothenburg, Sweden}
\affiliation{Yale University, New Haven, CT, 06520, USA}

\author{Hendrik Hansen-Goos}
\affiliation{Institute of Planetary Research, German Aerospace Center (DLR), Rutherfordstra{\ss}e 2, 12489 Berlin, Germany
}

\author{J. S. Wettlaufer}
\affiliation{Yale University, New Haven, CT, 06520, USA}
\affiliation{Nordic Institute for Theoretical Physics (NORDITA), 10691 Stockholm, Sweden}

\author{L. A. Wilen}
\affiliation{Unilever Research and Development, Trumbull, CT, 06611, USA}
\affiliation{Yale University, New Haven, CT, 06520, USA}

\date{\today}

\begin{abstract}
We describe an optical scattering study of grain boundary premelting in water ice.  Ubiquitous long ranged attractive polarization forces act  to suppress grain boundary melting whereas repulsive forces originating in screened Coulomb interactions and classical colligative effects enhance it.  The liquid enhancing effects can be manipulated by adding dopant ions to the system.  For all measured grain boundaries this leads to increasing premelted film thickness with increasing electrolyte concentration.  Although we understand that the interfacial surface charge densities $q_s$ and solute concentrations can potentially dominate the film thickness, we can not directly measure them within a given grain boundary.  Therefore, as a framework for interpreting the data we consider two appropriate $q_s$ dependent limits; one is dominated by the colligative effect and one is dominated by electrostatic interactions.

\end{abstract}


\maketitle

\section{Introduction}

Surface and interfacial premelting have been studied for the principal facets of ice and for ice crystals in contact with a wide range of other materials \cite{Dash:2006}.  Despite the ubiquity of polycrystalline materials in laboratory and natural environmental settings, and the basic relevance of grain boundaries in effective medium properties, direct measurements of grain boundary (GB) premelting for systems in thermodynamic equilibrium have not been made.  Crystallography insures that the mismatch between two grains is characterized by molecular scale disorder, the structure of which, as the melting temperature is approached from below, is basic to the edifice of premelting. However, in general liquid-like grain boundary structure is a controversial topic with a complex relationship between film thickness, temperature and chemical composition \cite[see e.g.,][]{Luo:2008, French:2010}.  For example, differing interpretations arise from studies in metals \cite{Glicksman1972,Hsieh1989}, colloidal crystals \cite{Alsayed2005}, and molecular solids such as benzene \cite{Craven1990}. Simulations suggest complete \cite{Broughton1986} and partial melting \cite{Mellenthin2008}, or the formation of a third orientation of the solid phase \cite{Ciccotti1983}.  A similar diversity of behavior is observed between grains in multicomponent systems \cite{Luo:2008, Shi2009, Shi2010}.  

\begin{figure}
\centering
\includegraphics[width=1.0\columnwidth, clip=true]{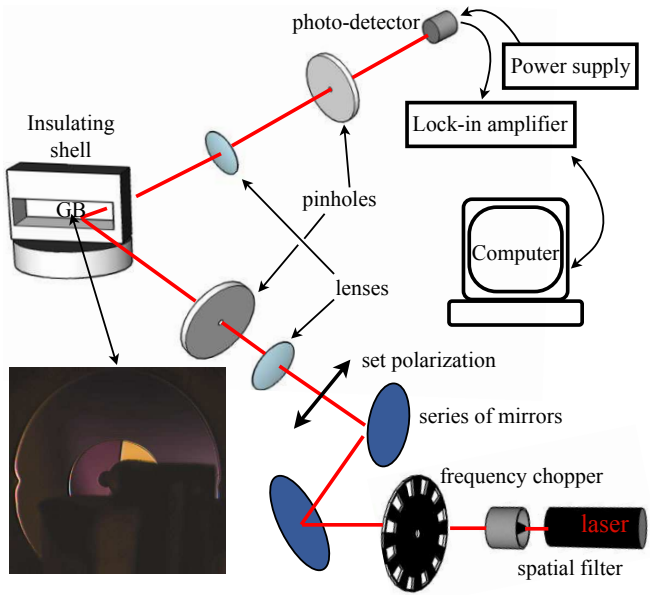}
\caption{Schematic of the optical bench geometry as described in \cite{Thomson2009b}.  A laser is spatially filtered and frequency chopped before the polarization is set and the beam is focused onto the grain boundary (GB).  The reflected signal is refocused onto a calibrated photo detector and data is output to a PC at selected time intervals.}
\label{fig:bench}
\end{figure}

The study of ice is compelling for many reasons \cite{Dash:2006}, a few of which we mention here.  First, it exhibits the same class of phase transitions found in more simply bound matter.  Second,  it can be held near the pure bulk melting point $T_m$ with relative ease.  Third, the results are of  immediate relevance to geophysical phenomena.  In order to probe the grain boundary between two ice crystals 
a light scattering apparatus was constructed to nucleate ice bicrystals, control their growth and to expose them to varying levels of ions using dissolved salt \cite{Thomson2009b}. As thermodynamic parameters are varied, laser light is reflected from a single grain boundary and the intensity of the reflected signal is interpreted using a theory that  incorporates the optical anisotropy of the bounding crystals \cite{Thomson2009a}.   In that theory we analyzed the reflection and transmission of plane waves by an isotropic layer (water) sandwiched between two uniaxial (ice) crystals of arbitrary orientation.  The experimental geometry is depicted schematically in Fig. \ref{fig:bench} and here we briefly summarize the approach detailed in \citet{Thomson2009b}.  

\section{\label{sec:exp}Experimental Method}

\begin{figure}
\centering
\includegraphics[width=1.0\columnwidth, clip=true]{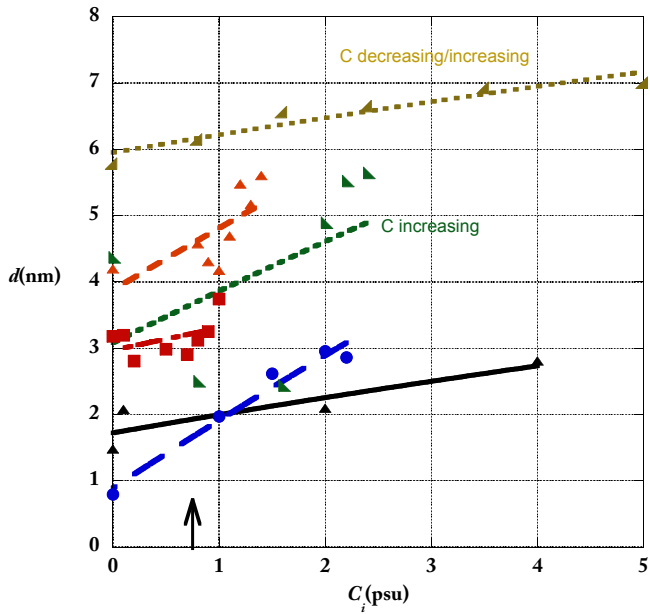}
\caption{Experimental grain boundary thickness $d$(nm) as a function of NaCl concentration $C_i$(psu) determined as described in the text.  Colors represent different grain boundary orientations determined from an independent measurement via polarimetry and a Schmidt plot as described in \cite{Thomson2009b}.  The orange and red data correspond to the grain boundaries with the largest lattice mismatch, and hence the largest magnitude of interfacial disorder at contact in the limit of zero impurity concentration.   The blue data show a small angle twist grain boundary and hence the weakest interfacial disorder at contact. Dark green (olive) points correspond to the same grain boundary as the concentration is cycled up (cycled down and back up) showing hysteresis. The lines are fits for the small $q_s$ colligative limit described by Eq. 
(\ref{eq:finald}). Finally, the vertical arrow corresponds to the upper bound of the surface charge density dependent critical composition above which the theory of \citep[][]{HansenGoos2010} predicts that $d$ increases with composition.  The upper and lower bounds depend on $q_s$, the former corresponding to $\sim$ 0.2 C m$^{-2}$ (the vertical arrow) and the latter to $\sim$ 0.02  C m$^{-2}$, which would be denoted by a vertical line on the origin.   As seen in Fig. \ref{fig:qsrange} when $q_s$ changes through this range so too does the compositional range of non-monotonic variations in $d$.  
The average colligatively shifted undercooling for these data is approximately 1.5 K.  The sources of error in the experimental points, with the maximum being 1.5 nm, are discussed in \S\ref{sec:exp}.}
\label{fig:Data}
\end{figure}

Our thermally insulated ice growth apparatus is used to nucleate and grow ice in contact with a solution of monovalent electrolyte impurity (NaCl).  The apparatus is mounted onto rotation and translation stages and placed in the path of a laser beam on an optical bench.  The incident beam is spatially filtered, frequency chopped, and polarized before it is focused onto the grain boundary and the reflected signal collected and analyzed.  In order to limit absorption and scattering within the ice, use of a low power 2.3 mW, 632.8 nm He-Ne laser insured that less than $0.02\%$ of the beam was attenuated within the ice itself. Hence, any heating effect was insignificant and below the resolution of our thermometry.  The reflected signal is refocused onto a photovoltaic detector whose output goes to a lock-in amplifier (SR830), and frequency-locks the measured signal to the incident beam.  The frequency-locked signal and temperature data are read by a computer at specified time intervals.  Bicrystals, with different crystallographic mismatch, are grown and we continuously collected the reflected intensity data as a function of the temperature and bulk impurity concentration.    From the measured voltage at the lock-in amplifier an intensity ratio is calculated.  The corrections include (a) accounting for the response of the photo detector, (b) the measurement algorithm of the lock-in amplifier, and (c) attenuation of the beam intensity by the optical elements of the system. Finally, the intensity ratio is converted to a film thickness by averaging many measurements taken at fixed bulk composition and temperature, using the theoretical model described in \citet{Thomson2009a}.  Consistent with other measurements \cite{Klein2004, Raviv2004} it  is assumed that the index of refraction of the film is the same as bulk water ($n_w=1.33$) which has weak thermal \cite{Saubade1981} and solutal \cite{CRC1976} dependence. The thickness values $d$  resulting from the measured intensity ratios are shown in Fig.~\ref{fig:Data}.  These data have statistically independent uncertainties due to the error in the measurement and intensity conversion ($\pm$ 3\%) and larger systematic uncertainties that result principally from the orientations of the crystals relative to the incident beam.  This systematic error is due to the coupling between the ice anisotropy and beam polarization, which can introduce constructive and destructive intensity fluctuations \cite{Thomson2009a}.  A quantitative error estimate of this latter uncertainty ranges from 5\%-25\%, increasing with increasing grain boundary thickness \cite{Thomson2010b}. This error is systematic because with the same experimental protocol and a given GB all points will be equally and unidirectionally displaced, thereby preserving trends with impurity concentration.  Therefore, we independently measure the crystal orientations and then determine GB thicknesses as a function of temperature and impurity concentration from the reflected light intensity measurements. 

In Fig.~\ref{fig:Data} the grain boundary thicknesses $d$ calculated as described above are shown for a range of experimental salt concentrations $C_i$ of the bulk liquid in contact with our ice grain boundaries.  The analytical precision of our $C_i$ measurements is $\pm 0.1$ psu,  and in the $C_i$ range from 0 to 5 psu, we observed grain boundary thicknesses from approximately 1 to 8 nm.  We have no direct measurements of the composition {\em within} the GBs, although the experimental waiting times are commensurate with the equilibration of a {\em bulk} salt solution via solute diffusion.  Therefore, as described in \S\ref{sec:theory}, we consider two limits of solutal equilibration when interpreting the experimental data.  When $C_i$ = 0 psu, a finite GB thickness is observed, associated with different values for each crystallographic misorientation, and for each GB the thickness increases with $C_i$. We discuss in detail below the interpretation of the intergrain boundary variation at lower solute concentrations.  

\section{\label{sec:theory}Theory}

A theoretical treatment of the role of impurities in interfacial premelting aides in the interpretation of the data.  The approach predicts that in the presence of soluble electrolyte impurities GB premelting in ice may be macroscopic very near the bulk melting temperature \cite{Benatov2004, Thomson2010a}.  A GB liquid film of thickness $d$ arises from the competition between attractive van der Waals interactions, colligative effects and repulsive interactions of electrostatic origin;  the presence of a surface charge $q_s$ is mediated by ions within the liquid layer.   Minimizing a free energy that combines all of these effects \cite{Wettlaufer:1999, Benatov2004, HansenGoos2010} provides the equilibrium film thickness $d$ at an impure grain boundary held at a temperature $T$ as follows; 
\begin{equation}
\rho_l q_m\frac{\Delta T}{T_m}=\frac{R_gT_mN_i}{d} -\frac{A_H}{6 \pi d^3}+\frac{q_s^2}{\epsilon \epsilon_o}\left[1-\frac{1}{\kappa d}\right] e^{-\kappa d}. 
\label{eq:dtherm} 
\end{equation}
Here, $\rho_l$ and  $q_m$ are the liquid density and the latent heat of fusion, $\Delta T=T_m-T$ is the undercooling relative to the melting temperature of a pure ice-water system,  $N_i$ is the number of moles of electrolyte per unit area, $A_H$ is the Hamaker constant, $\epsilon$ and $\epsilon_o$ are the relative and free space permittivities, $\kappa^{-1} =\sqrt{\frac{\epsilon\epsilon_0 k_B T d}{\text{e}^2 N_A N_i}}$ is the Debye length, and $R_g$, $k_B$, $N_A$ and $\text{e}$ are the ideal gas and Boltzmann constants, Avogadro's number and the elementary charge respectively.  Although in general the frequency dependence of the dielectric response of ice/water systems can lead to important retarded potential effects \cite{Dash:2006} that can in principle influence a grain boundary film \cite{Benatov2004}, such effects are presently too subtle to be probed.  More importantly, the Hamaker constant of relevance in Eq.~(\ref{eq:dtherm}) is $A_H=3.3\times10^{-22}$, which is calculated from full frequency dependent dispersion force theory \citep[see e.g.,][]{HansenGoos2010}, and hence renders this attractive--film suppressing--term negligible for the range of film thicknesses observed experimentally. The question here is what limits of Eq. \ref{eq:dtherm} can explain the experimental data?  
There are two limits of the theory that are relevant to consider.

We take one limit where $q_s$ is very small, and there is very slow equilibration of composition along the GB.  Here, the effect is principally colligative and hence $d$ increases with $C_i$.  The small $q_s$ limit of Eq. (\ref{eq:dtherm}) and vanishing dispersion forces is represented by Eq. (29) of \citet[][]{HansenGoos2010}.  As shown in Fig.~\ref{fig:qsrange}, the theory applies to the case when the solute along the grain boundary (or grain boundary network in bulk ice) is isolated and the absolute undercooling is specified. The total number of moles, and hence areal concentration, $N_i$, of solute is conserved. This is relevant, and has been applied, to glaciers and polar ice where the diffusion of impurities is slow enough that the grain boundaries in the interior may be considered to be isolated.

In the second limit $q_s$ plays a larger role and we fit its dependence on $C_i$ in a manner consistent with Eq.~\eqref{eq:dtherm}.  Here, $d$ 
also increases with $C_i$  in the GB,  which is equilibrated to the bulk fluid reservoir.  The dominating effect is the screened Coulomb repulsion between the two sides of the GB.  We discuss these limits in turn. 

\subsection{Small $q_s$, Slow Equilibration, Locally Colligative Limit}

A motivation for this limit is that data are qualitatively consistent with an assumption that the actual concentration $N_i$ along the grain boundary is proportional to the bulk concentration of the reservoir into which the ice grows. Thus, one may write $N_i = C_i  \ell_{\text{eff}}$ where $\ell_{\text{eff}}$ is a length scale that relates the two quantities. Physically, one can think of this length scale as being related to a ``funneling'' effect from the grain boundary groove as the ice lens is grown into the bulk solution, trapping the total solute across this effective length in the grain boundary. Hence we are then led to consider the possibility that, on the time scale over which the measurement is performed, the bulk concentration in the GB remains greater than that of the reservoir either (a) due to slow diffusion, possibly influenced by finite size effects, or (b) due to effects that lower the chemical potential along the grain boundary in a way not captured in the current theory, but along the lines discussed in \cite[][]{Rempel:2001}.

\begin{figure}
\centering
\includegraphics[width=1.0\columnwidth, clip=true]{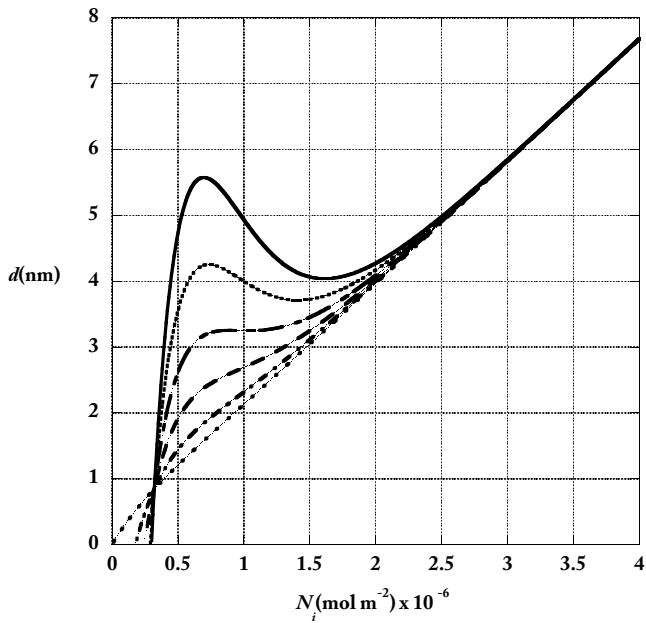}
\caption{The theoretical dependence of the film thickness $d$ on impurity level $N_i$ in moles of electrolyte per unit area (here written as mol m$^{-2}$) for the range of  surface charge density $q_s$ from $0.02$ to $0.22$ C m$^{-2}$.  The undercooling $\Delta T$ is unity, which is the approximate average experimental value. The surface charge $q_s$ increases from $0.02$ in the lowest curve to $0.22$ in the upper solid curve in increments of 0.04 C m$^{-2}$ and the calculation becomes meaningless below a cut off of approximately the molecular scale.  The essential points are (a) that $d$ depends sensitively on $q_s$ at very low dopant levels, (b) for $N_i > N_{\text{th}} \simeq 8q_s ~\mu$mol m$^{-2}$, $d$ will always increase with $N_i$ (for any undercooling) and (c) over a range of undercooling there is a range of $q_s$ at small impurity concentrations where $d$ {\em decreases} as $N_i$ increases, as discussed in  \citet[][]{HansenGoos2010}.}
\label{fig:qsrange}
\end{figure}

This limit is plotted in Fig.~\ref{fig:qsrange} where we see behavior similar to the experimental fluctuations at low $N_i$ and the experimental trend of increasing film thickness with the areal impurity level.  There is a sensitive dependence of the film thickness on $q_s$ at very low dopant levels, where the slope changes sign below a threshold value $N_{\text{th}} \simeq 8q_s ~\mu$mol m$^{-2}$.  However, when $N_i > N_{\text{th}}$ the value of $d$ always increases with $N_i$.

To be more quantitative, if we postulate the above, then we can express the colligative limit of Eq. \eqref{eq:dtherm} as follows:
\begin{equation}
\frac{\rho_l q_m}{T_m}{\Delta T} = \frac{\rho_l q_m}{T_m}\left[\Delta T^\prime + \frac{R_g {T_m}^2 C_i}{\rho_l q_m} \right]= R_g T_m \frac{C_i \ell_{\text{eff}}}{d}\, ,
\end{equation}
where $\Delta T^{\prime}$ is the undercooling with respect to the solutally depressed melting temperature. 
Upon rearranging, we find
\begin{equation}
d = \ell_{\text{eff}} \left[1 + \frac{\rho_l q_m \Delta T^\prime}{ R_g {T_m}^2 C_i} \right]^{-1}. 
\end{equation}
Finally, we further assume that there is a minimum thickness $d_\text{o}$ to the grain boundary due to short range effects and hence we can write
\begin{equation}
d = d_\text{o} + \ell_{\text{eff}} \left[1 + \frac{\rho_l q_m \Delta T^\prime}{ R_g {T_m}^2 C_i} \right]^{-1},  
\label{eq:finald} 
\end{equation}
where $d_\text{o}$ will depend on the particular grain boundary mismatch.  The straight lines in Fig.~\ref{fig:Data} are the fits using Eq.~\eqref{eq:finald} for which we find an average $\ell_{\text{eff}}$ of approximately 30 nm and used the average $\Delta T^\prime$ for each experiment.  

\subsection{Large $q_s$,  Fast Equilibration, Electrostatic Limit}

\begin{figure}
\centering
\includegraphics[width=1.0\columnwidth, clip=true]{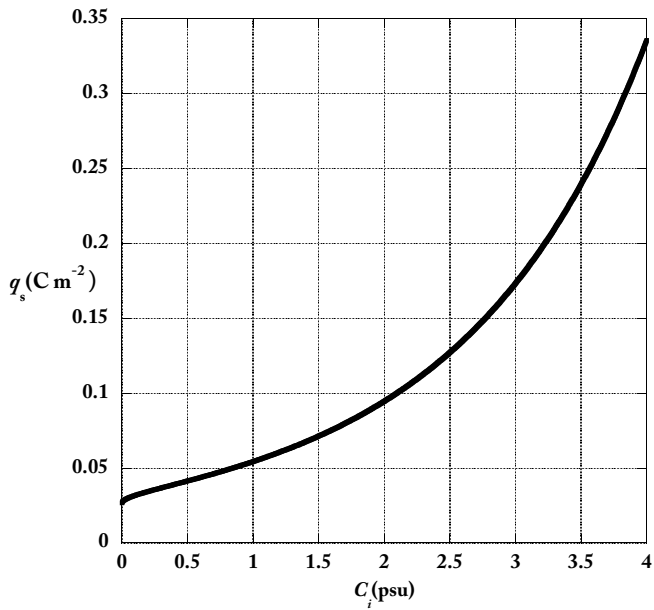}
\caption{The surface charge density $q_s$ as a function of the concentration $C_i$ of the bulk fluid reservoir assumed to be in equilibrium with the GB. }
\label{fig:qs}
\end{figure}

\begin{figure}
\centering
\includegraphics[width=1.0\columnwidth, clip=true]{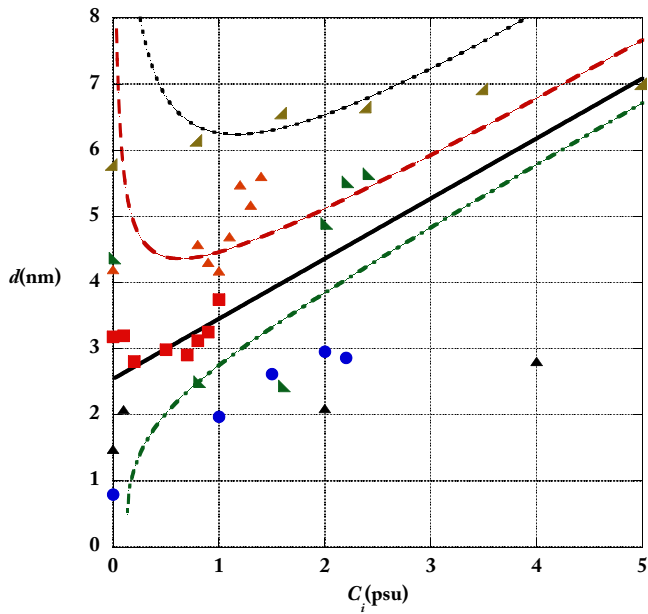}
\caption{Theoretical prediction of grain boundary thickness $d$(nm) as a function of $C_i$(psu) determined as described in the text surrounding Eq.~\eqref{eq:prime}  for the large $q_s$, electrostatic limit.  From the lower line to the upper line the undercoolings are 2.0, 1.5, 1.0 and 0.5 K.  The data are replotted with the same symbols as in Fig.~\ref{fig:Data} but here to avoid confusion we leave off the fits from Eq.~\eqref{eq:finald}.  We note that the data are all ostensibly at the same undercooling of about 1.5K.  It is not our intent to fit the data from any given GB in this figure, but simply to show the overall behavior of this limit of the theory and the range over which $d$ increases with $C_i$. The sources of error in the experimental points, with the maximum being 1.5 nm, are discussed in \S\ref{sec:exp}.}
\label{fig:lim2}
\end{figure}

An experimental motivation for this limit is that the measured GB always extends to the bulk fluid reservoir outside the ice, and the temperature is measured relative to that at the ice-solution edge.  Thus, we consider the case when the solute in the GB is diffusionally connected to an external reservoir of specified bulk solute concentration. In equilibrium, then, one might expect the volume concentration of solute in the grain boundary to be equal to that of the reservoir. In this limit, one may replace $N_i/d$ by  $C_i$ in Eq.~\eqref{eq:dtherm}.  Rearranging terms yields 
\begin{equation}
\rho_l q_m\frac{\Delta T^{\prime}}{T_m} = \frac{q_s^2}{\epsilon \epsilon_o}\left[1-\frac{1}{\kappa d}\right] e^{-\kappa d} -\frac{A_H}{6 \pi d^3} \, .
\label{eq:prime} 
\end{equation}
As already noted, we have no experimental measurements of the interfacial surface charge and we have no direct measure of the composition within the GB.  Thus, we take as an ansatz that $q_s$ is a function of  the concentration $C_i$. 
The experimental data shown in Figs. \ref{fig:Data} and \ref{fig:lim2} demonstrate a substantial scatter but nonetheless motivate one to consider a linear dependence of the grain boundary thickness $d$ on the impurity concentration $C_i$.  Thus, for an undercooling of $1.5\text{ K}$ we take $d=3\text{ nm}$ for $C_i=0.5\text{psu}$ and $d=4\text{ nm}$ for $C_i=1.6\text{psu}$ and linearly extrapolate. Now, neglecting the temperature dependence of $q_s$ over the relatively small experimental temperature ranges,  we use Eq.~\eqref{eq:prime}  to calculate $q_s(C_i)$, the result of which is shown in Fig.~\ref{fig:qs}, and we use this in Eq.~\eqref{eq:prime} to determine $d$ as a function of $C_i$ for undercoolings from 0.5 to 2.0 K.  We compare these theoretical results with the experimental data in Fig.~\ref{fig:lim2}, in which we see that beyond  $C_i \approx 1 \text{psu}$
the general linear increase in the GB thickness with concentration is captured.  The inter-experimental variability at low values of $C_i$ is again seen here;  depending on the undercooling $d$ can either increase {\em or} decrease with  $C_i$. 

\section{Discussion and Conclusion}

We do not have a testable microscopic description relating the measured crystallographic orientations of the ice crystals defining the GB to the nature of the disorder at short range.  Such a theory may facilitate some interpretation of the variation of the observed intercepts $d_\text{o}$ in Fig. \ref{fig:Data}, but we note that they range from about 4 to 24 molecular layers, and bulk fluid properties have been demonstrated in subnanometer scale water films \cite{Klein2004, Raviv2004}.     As seen in Figs. \ref{fig:qsrange} and \ref{fig:lim2} for a wide range of values of the surface charge density, the regions  with $N_i < N_{\text{th}}$ exhibit the largest variation in film thickness with $q_s$.  Hence, in this context we interpret the observed fluctuations in $d \sim d_\text{o}$ as being associated with the combined sensitivity of the magnitude of $d$ on $q_s$ and of the sign and magnitude of the slope of  the film thickness with impurity level.   

Each GB is grown with different thermal and solutal histories, although the intergranular ions can exchange and relax with the bulk reservoir, the fluctuations associated with these histories leave unique ionic concentrations behind, bound to differently charged surfaces.  As described above, it is evident that the effects associated with additional sensitivities when $N_i \lesssim N_{\text{th}}$ can amplify such history dependence at low dopant levels.  Because we do not have intra-GB film measurements of ion concentration, to speculate further than is warranted by direct evidence risks interpretation that cannot be tested quantitatively with our method. 

These measurements provide essential  information for the effective medium properties of ice polycrystals.  For example, the interfaces between grains provide a ready pathway for the transport of soluble impurities and isotopes, which in the case of Earth's ice sheets act as high resolution proxies for past climates.  Indeed, it has been shown that the liquid veins where three grains abut and the nodes where four or more grains terminate can act as the principal conduits for transport of such proxies \cite[e.g.,][]{Rempel:2001}.  However, we know that the total intergrain surface area is dominated by grain boundaries; that is the faces between two grains.  Therefore, while we have found that premelted films at grain boundaries are 1-10 nm thick, they may dominate the volume of liquid through which transport is controlled.  Additional relevant settings include how such interfacial liquidity influences the rates of atmospheric chemical processes, which take place on polycrystalline ice particles, the electrical conductivity of glaciers and ice sheets, frost heave, thunderstorm electrification and extraterrestrial ices \cite{Dash:2006}. 

In summary we have described an optical scattering study of grain boundary premelting in ice.  By doping the grain boundaries with ions to provide a colligative effect, and a source of repulsive screened Coulomb interactions, we found that under the experimental conditions such liquid enhancing effects dominate liquid suppressing long ranged attractive dispersion forces.  In all grain boundaries we find that the premelted intergranular film thickness increased with electrolyte concentration and at the lowest concentrations the analysis demonstrated substantial sensitivity to the surface charge density at the grain surfaces.  Finally, the finding of films of thicknesses up to about 10 nm has a range of immediate environmental applications.

\begin{acknowledgments}

EST, HH-G, and JSW thank the Helmholtz Gemeinschaft Alliance ``Planetary Evolution and Life''.  EST thanks the University of Gothenburg, and the Nordic Top-Level Research Initiative CRAICC for support. JSW thanks the Wenner-Gren Foundation, the John Simon Guggenheim Foundation, and the Swedish Research Council for generous support of this research.

\end{acknowledgments}


\begin{thebibliography}{23}%
\makeatletter
\providecommand \@ifxundefined [1]{%
 \@ifx{#1\undefined}
}%
\providecommand \@ifnum [1]{%
 \ifnum #1\expandafter \@firstoftwo
 \else \expandafter \@secondoftwo
 \fi
}%
\providecommand \@ifx [1]{%
 \ifx #1\expandafter \@firstoftwo
 \else \expandafter \@secondoftwo
 \fi
}%
\providecommand \natexlab [1]{#1}%
\providecommand \enquote  [1]{``#1''}%
\providecommand \bibnamefont  [1]{#1}%
\providecommand \bibfnamefont [1]{#1}%
\providecommand \citenamefont [1]{#1}%
\providecommand \href@noop [0]{\@secondoftwo}%
\providecommand \href [0]{\begingroup \@sanitize@url \@href}%
\providecommand \@href[1]{\@@startlink{#1}\@@href}%
\providecommand \@@href[1]{\endgroup#1\@@endlink}%
\providecommand \@sanitize@url [0]{\catcode `\\12\catcode `\$12\catcode
  `\&12\catcode `\#12\catcode `\^12\catcode `\_12\catcode `\%12\relax}%
\providecommand \@@startlink[1]{}%
\providecommand \@@endlink[0]{}%
\providecommand \url  [0]{\begingroup\@sanitize@url \@url }%
\providecommand \@url [1]{\endgroup\@href {#1}{\urlprefix }}%
\providecommand \urlprefix  [0]{URL }%
\providecommand \Eprint [0]{\href }%
\providecommand \doibase [0]{http://dx.doi.org/}%
\providecommand \selectlanguage [0]{\@gobble}%
\providecommand \bibinfo  [0]{\@secondoftwo}%
\providecommand \bibfield  [0]{\@secondoftwo}%
\providecommand \translation [1]{[#1]}%
\providecommand \BibitemOpen [0]{}%
\providecommand \bibitemStop [0]{}%
\providecommand \bibitemNoStop [0]{.\EOS\space}%
\providecommand \EOS [0]{\spacefactor3000\relax}%
\providecommand \BibitemShut  [1]{\csname bibitem#1\endcsname}%
\let\auto@bib@innerbib\@empty
\bibitem [{\citenamefont {Dash}, \citenamefont {Rempel},\ and\ \citenamefont
  {Wettlaufer}(2006)}]{Dash:2006}%
  \BibitemOpen
  \bibfield  {author} {\bibinfo {author} {\bibfnamefont {J.~G.}\ \bibnamefont
  {Dash}}, \bibinfo {author} {\bibfnamefont {A.~W.}\ \bibnamefont {Rempel}}, \
  and\ \bibinfo {author} {\bibfnamefont {J.~S.}\ \bibnamefont {Wettlaufer}},\
  }\href {\doibase DOI 10.1103/RevModPhys.78.695} {\bibfield  {journal}
  {\bibinfo  {journal} {Rev. Mod. Phys.}\ }\textbf {\bibinfo {volume} {78}},\
  \bibinfo {pages} {695} (\bibinfo {year} {2006})}\BibitemShut {NoStop}%
\bibitem [{\citenamefont {Luo}\ and\ \citenamefont {Chiang}(2008)}]{Luo:2008}%
  \BibitemOpen
  \bibfield  {author} {\bibinfo {author} {\bibfnamefont {J.}~\bibnamefont
  {Luo}}\ and\ \bibinfo {author} {\bibfnamefont {Y.-M.}\ \bibnamefont
  {Chiang}},\ }\href {\doibase DOI 10.1146/annurev.matsci.38.060407.132431}
  {\bibfield  {journal} {\bibinfo  {journal} {Annu. Rev. Mater. Res.}\ }\textbf
  {\bibinfo {volume} {38}},\ \bibinfo {pages} {227} (\bibinfo {year}
  {2008})}\BibitemShut {NoStop}%
\bibitem [{\citenamefont {French}\ \emph {et~al.}(2010)\citenamefont {French},
  \citenamefont {Parsegian}, \citenamefont {Podgornik}, \citenamefont {Rajter},
  \citenamefont {Jagota}, \citenamefont {Luo}, \citenamefont {Asthagiri},
  \citenamefont {Chaudhury}, \citenamefont {Chiang}, \citenamefont {Granick},
  \citenamefont {Kalinin}, \citenamefont {Kardar}, \citenamefont {Kjellander},
  \citenamefont {Langreth}, \citenamefont {Lewis}, \citenamefont {Lustig},
  \citenamefont {Wesolowski}, \citenamefont {Wettlaufer}, \citenamefont
  {Ching}, \citenamefont {Finnis}, \citenamefont {Houlihan}, \citenamefont {von
  Lilienfeld}, \citenamefont {van Oss},\ and\ \citenamefont
  {Zemb}}]{French:2010}%
  \BibitemOpen
  \bibfield  {author} {\bibinfo {author} {\bibfnamefont {R.~H.}\ \bibnamefont
  {French}}, \bibinfo {author} {\bibfnamefont {V.~A.}\ \bibnamefont
  {Parsegian}}, \bibinfo {author} {\bibfnamefont {R.}~\bibnamefont
  {Podgornik}}, \bibinfo {author} {\bibfnamefont {R.~F.}\ \bibnamefont
  {Rajter}}, \bibinfo {author} {\bibfnamefont {A.}~\bibnamefont {Jagota}},
  \bibinfo {author} {\bibfnamefont {J.}~\bibnamefont {Luo}}, \bibinfo {author}
  {\bibfnamefont {D.}~\bibnamefont {Asthagiri}}, \bibinfo {author}
  {\bibfnamefont {M.~K.}\ \bibnamefont {Chaudhury}}, \bibinfo {author}
  {\bibfnamefont {Y.-m.}\ \bibnamefont {Chiang}}, \bibinfo {author}
  {\bibfnamefont {S.}~\bibnamefont {Granick}}, \bibinfo {author} {\bibfnamefont
  {S.}~\bibnamefont {Kalinin}}, \bibinfo {author} {\bibfnamefont
  {M.}~\bibnamefont {Kardar}}, \bibinfo {author} {\bibfnamefont
  {R.}~\bibnamefont {Kjellander}}, \bibinfo {author} {\bibfnamefont {D.~C.}\
  \bibnamefont {Langreth}}, \bibinfo {author} {\bibfnamefont {J.}~\bibnamefont
  {Lewis}}, \bibinfo {author} {\bibfnamefont {S.}~\bibnamefont {Lustig}},
  \bibinfo {author} {\bibfnamefont {D.}~\bibnamefont {Wesolowski}}, \bibinfo
  {author} {\bibfnamefont {J.~S.}\ \bibnamefont {Wettlaufer}}, \bibinfo
  {author} {\bibfnamefont {W.-Y.}\ \bibnamefont {Ching}}, \bibinfo {author}
  {\bibfnamefont {M.}~\bibnamefont {Finnis}}, \bibinfo {author} {\bibfnamefont
  {F.}~\bibnamefont {Houlihan}}, \bibinfo {author} {\bibfnamefont {O.~A.}\
  \bibnamefont {von Lilienfeld}}, \bibinfo {author} {\bibfnamefont {C.~J.}\
  \bibnamefont {van Oss}}, \ and\ \bibinfo {author} {\bibfnamefont
  {T.}~\bibnamefont {Zemb}},\ }\href {\doibase DOI 10.1103/RevModPhys.82.1887}
  {\bibfield  {journal} {\bibinfo  {journal} {Rev. Mod. Phys.}\ }\textbf
  {\bibinfo {volume} {82}},\ \bibinfo {pages} {1887} (\bibinfo {year}
  {2010})}\BibitemShut {NoStop}%
\bibitem [{\citenamefont {Glicksman}\ and\ \citenamefont
  {Vold}(1972)}]{Glicksman1972}%
  \BibitemOpen
  \bibfield  {author} {\bibinfo {author} {\bibfnamefont {M.~E.}\ \bibnamefont
  {Glicksman}}\ and\ \bibinfo {author} {\bibfnamefont {C.~L.}\ \bibnamefont
  {Vold}},\ }\href@noop {} {\bibfield  {journal} {\bibinfo  {journal} {Surf.
  Sci.}\ }\textbf {\bibinfo {volume} {31}},\ \bibinfo {pages} {50} (\bibinfo
  {year} {1972})}\BibitemShut {NoStop}%
\bibitem [{\citenamefont {Hsieh}\ and\ \citenamefont
  {Balluffi}(1989)}]{Hsieh1989}%
  \BibitemOpen
  \bibfield  {author} {\bibinfo {author} {\bibfnamefont {T.~E.}\ \bibnamefont
  {Hsieh}}\ and\ \bibinfo {author} {\bibfnamefont {R.~W.}\ \bibnamefont
  {Balluffi}},\ }\href@noop {} {\bibfield  {journal} {\bibinfo  {journal} {Acta
  Metall.}\ }\textbf {\bibinfo {volume} {37}},\ \bibinfo {pages} {1637}
  (\bibinfo {year} {1989})}\BibitemShut {NoStop}%
\bibitem [{\citenamefont {Alsayed}\ \emph {et~al.}(2005)\citenamefont
  {Alsayed}, \citenamefont {Islam}, \citenamefont {Zhang}, \citenamefont
  {Collings},\ and\ \citenamefont {Yodh}}]{Alsayed2005}%
  \BibitemOpen
  \bibfield  {author} {\bibinfo {author} {\bibfnamefont {A.}~\bibnamefont
  {Alsayed}}, \bibinfo {author} {\bibfnamefont {M.}~\bibnamefont {Islam}},
  \bibinfo {author} {\bibfnamefont {J.}~\bibnamefont {Zhang}}, \bibinfo
  {author} {\bibfnamefont {P.}~\bibnamefont {Collings}}, \ and\ \bibinfo
  {author} {\bibfnamefont {A.}~\bibnamefont {Yodh}},\ }\href@noop {} {\bibfield
   {journal} {\bibinfo  {journal} {Science}\ }\textbf {\bibinfo {volume} {309}}
  (\bibinfo {year} {2005})}\BibitemShut {NoStop}%
\bibitem [{\citenamefont {Craven}\ \emph {et~al.}(1990)\citenamefont {Craven},
  \citenamefont {Dosseh}, \citenamefont {Rousseau},\ and\ \citenamefont
  {Fuchs}}]{Craven1990}%
  \BibitemOpen
  \bibfield  {author} {\bibinfo {author} {\bibfnamefont {C.~J.}\ \bibnamefont
  {Craven}}, \bibinfo {author} {\bibfnamefont {G.}~\bibnamefont {Dosseh}},
  \bibinfo {author} {\bibfnamefont {B.}~\bibnamefont {Rousseau}}, \ and\
  \bibinfo {author} {\bibfnamefont {A.~H.}\ \bibnamefont {Fuchs}},\ }\href@noop
  {} {\bibfield  {journal} {\bibinfo  {journal} {J. Phys. (Paris)}\ }\textbf
  {\bibinfo {volume} {51}},\ \bibinfo {pages} {2489} (\bibinfo {year}
  {1990})}\BibitemShut {NoStop}%
\bibitem [{\citenamefont {Broughton}\ and\ \citenamefont
  {Gilmer}(1986)}]{Broughton1986}%
  \BibitemOpen
  \bibfield  {author} {\bibinfo {author} {\bibfnamefont {J.~Q.}\ \bibnamefont
  {Broughton}}\ and\ \bibinfo {author} {\bibfnamefont {G.~H.}\ \bibnamefont
  {Gilmer}},\ }\href@noop {} {\bibfield  {journal} {\bibinfo  {journal} {Phys.
  Rev. Lett.}\ }\textbf {\bibinfo {volume} {56}},\ \bibinfo {pages} {2692}
  (\bibinfo {year} {1986})}\BibitemShut {NoStop}%
\bibitem [{\citenamefont {Mellenthin}, \citenamefont {Karma},\ and\
  \citenamefont {Plapp}(2008)}]{Mellenthin2008}%
  \BibitemOpen
  \bibfield  {author} {\bibinfo {author} {\bibfnamefont {J.}~\bibnamefont
  {Mellenthin}}, \bibinfo {author} {\bibfnamefont {A.}~\bibnamefont {Karma}}, \
  and\ \bibinfo {author} {\bibfnamefont {M.}~\bibnamefont {Plapp}},\
  }\href@noop {} {\bibfield  {journal} {\bibinfo  {journal} {Phys. Rev. B}\
  }\textbf {\bibinfo {volume} {78}} (\bibinfo {year} {2008})}\BibitemShut
  {NoStop}%
\bibitem [{\citenamefont {Ciccotti}, \citenamefont {Guillope},\ and\
  \citenamefont {Pontikis}(1983)}]{Ciccotti1983}%
  \BibitemOpen
  \bibfield  {author} {\bibinfo {author} {\bibfnamefont {G.}~\bibnamefont
  {Ciccotti}}, \bibinfo {author} {\bibfnamefont {M.}~\bibnamefont {Guillope}},
  \ and\ \bibinfo {author} {\bibfnamefont {V.}~\bibnamefont {Pontikis}},\
  }\href@noop {} {\bibfield  {journal} {\bibinfo  {journal} {Phys. Rev. B}\
  }\textbf {\bibinfo {volume} {27}},\ \bibinfo {pages} {5576} (\bibinfo {year}
  {1983})}\BibitemShut {NoStop}%
\bibitem [{\citenamefont {Shi}\ and\ \citenamefont {Luo}(2009)}]{Shi2009}%
  \BibitemOpen
  \bibfield  {author} {\bibinfo {author} {\bibfnamefont {X.}~\bibnamefont
  {Shi}}\ and\ \bibinfo {author} {\bibfnamefont {J.}~\bibnamefont {Luo}},\
  }\href {\doibase DOI 10.1063/1.3155443} {\bibfield  {journal} {\bibinfo
  {journal} {Appl. Phys. Lett.}\ }\textbf {\bibinfo {volume} {94}},\ \bibinfo
  {pages} {251908} (\bibinfo {year} {2009})}\BibitemShut {NoStop}%
\bibitem [{\citenamefont {Shi}\ and\ \citenamefont {Luo}(2010)}]{Shi2010}%
  \BibitemOpen
  \bibfield  {author} {\bibinfo {author} {\bibfnamefont {X.}~\bibnamefont
  {Shi}}\ and\ \bibinfo {author} {\bibfnamefont {J.}~\bibnamefont {Luo}},\
  }\href {\doibase 10.1103/PhysRevLett.105.236102} {\bibfield  {journal}
  {\bibinfo  {journal} {Phys. Rev. Lett.}\ }\textbf {\bibinfo {volume} {105}},\
  \bibinfo {pages} {236102} (\bibinfo {year} {2010})}\BibitemShut {NoStop}%
\bibitem [{\citenamefont {Thomson}, \citenamefont {Wettlaufer},\ and\
  \citenamefont {Wilen}(2009)}]{Thomson2009b}%
  \BibitemOpen
  \bibfield  {author} {\bibinfo {author} {\bibfnamefont {E.~S.}\ \bibnamefont
  {Thomson}}, \bibinfo {author} {\bibfnamefont {J.~S.}\ \bibnamefont
  {Wettlaufer}}, \ and\ \bibinfo {author} {\bibfnamefont {L.~A.}\ \bibnamefont
  {Wilen}},\ }\href {\doibase 10.1063/1.3249562} {\bibfield  {journal}
  {\bibinfo  {journal} {Rev. Sci. Instrum.}\ }\textbf {\bibinfo {volume}
  {80}},\ \bibinfo {eid} {103903} (\bibinfo {year} {2009})}\BibitemShut
  {NoStop}%
\bibitem [{\citenamefont {Thomson}, \citenamefont {Wilen},\ and\ \citenamefont
  {Wettlaufer}(2009)}]{Thomson2009a}%
  \BibitemOpen
  \bibfield  {author} {\bibinfo {author} {\bibfnamefont {E.~S.}\ \bibnamefont
  {Thomson}}, \bibinfo {author} {\bibfnamefont {L.~A.}\ \bibnamefont {Wilen}},
  \ and\ \bibinfo {author} {\bibfnamefont {J.~S.}\ \bibnamefont {Wettlaufer}},\
  }\href@noop {} {\bibfield  {journal} {\bibinfo  {journal} {J. Phys. Condens.
  Matter}\ }\textbf {\bibinfo {volume} {21}},\ \bibinfo {pages} {195407}
  (\bibinfo {year} {2009})}\BibitemShut {NoStop}%

\bibitem [{\citenamefont {Klein}\ \emph {et~al.}(2004)\citenamefont {Klein},
  \citenamefont {Raviv}, \citenamefont {Perkin}, \citenamefont {Kampf},
  \citenamefont {Chai},\ and\ \citenamefont {Giasson}}]{Klein2004}%
  \BibitemOpen
  \bibfield  {author} {\bibinfo {author} {\bibfnamefont {J.}~\bibnamefont
  {Klein}}, \bibinfo {author} {\bibfnamefont {U.}~\bibnamefont {Raviv}},
  \bibinfo {author} {\bibfnamefont {S.}~\bibnamefont {Perkin}}, \bibinfo
  {author} {\bibfnamefont {N.}~\bibnamefont {Kampf}}, \bibinfo {author}
  {\bibfnamefont {L.}~\bibnamefont {Chai}}, \ and\ \bibinfo {author}
  {\bibfnamefont {S.}~\bibnamefont {Giasson}},\ }\href {\doibase DOI
  10.1088/0953-8984/16/45/008} {\bibfield  {journal} {\bibinfo  {journal} {J.
  Phys. Condens. Matter}\ }\textbf {\bibinfo {volume} {16}},\ \bibinfo {pages}
  {S5437} (\bibinfo {year} {2004})}\BibitemShut {NoStop}%
\bibitem [{\citenamefont {Raviv}\ \emph {et~al.}(2004)\citenamefont {Raviv},
  \citenamefont {Perkin}, \citenamefont {Laurat},\ and\ \citenamefont
  {Klein}}]{Raviv2004}%
  \BibitemOpen
  \bibfield  {author} {\bibinfo {author} {\bibfnamefont {U.}~\bibnamefont
  {Raviv}}, \bibinfo {author} {\bibfnamefont {S.}~\bibnamefont {Perkin}},
  \bibinfo {author} {\bibfnamefont {P.}~\bibnamefont {Laurat}}, \ and\ \bibinfo
  {author} {\bibfnamefont {J.}~\bibnamefont {Klein}},\ }\href {\doibase DOI
  10.1021/la030419d} {\bibfield  {journal} {\bibinfo  {journal} {Langmuir}\
  }\textbf {\bibinfo {volume} {20}},\ \bibinfo {pages} {5322} (\bibinfo {year}
  {2004})}\BibitemShut {NoStop}%
\bibitem [{\citenamefont {Saubade}(1981)}]{Saubade1981}%
  \BibitemOpen
  \bibfield  {author} {\bibinfo {author} {\bibfnamefont {C.}~\bibnamefont
  {Saubade}},\ }\href@noop {} {\bibfield  {journal} {\bibinfo  {journal} {J.
  Phys. (Paris)}\ }\textbf {\bibinfo {volume} {42}},\ \bibinfo {pages} {359}
  (\bibinfo {year} {1981})}\BibitemShut {NoStop}%
%
\bibitem [{\citenamefont {Weast}(1976)}]{CRC1976}%
  \BibitemOpen
  \bibinfo {editor} {\bibfnamefont {R.~C.}\ \bibnamefont {Weast}},\ ed.,\
  \enquote {\bibinfo {title} {Handbook of chemistry and physics},}\ \ (\bibinfo
   {publisher} {CRC Press},\ \bibinfo {address} {Cleveland, Ohio},\ \bibinfo
  {year} {1976})\ pp.\ \bibinfo {pages} {D--252 to D--253},\ \bibinfo {edition}
  {57th}\ ed.\BibitemShut {Stop}%
  %
 \bibitem [{\citenamefont {Thomson}(2010)}]{Thomson2010b}%
  \BibitemOpen
  \bibfield  {author} {\bibinfo {author} {\bibfnamefont {E.~S.}\ \bibnamefont
  {Thomson}},\ }\emph {\bibinfo {title} {An Optical Study of Ice Grain
  Boundaries}},\ \href@noop {} {Ph.D. thesis},\ \bibinfo  {school} {Yale
  University}, \bibinfo {address} {New Haven, CT, USA} (\bibinfo {year}
  {2010})\BibitemShut {NoStop}%
%
%
\bibitem [{\citenamefont {Hansen-Goos}\ and\ \citenamefont
  {Wettlaufer}(2010)}]{HansenGoos2010}%
  \BibitemOpen
  \bibfield  {author} {\bibinfo {author} {\bibfnamefont {H.}~\bibnamefont
  {Hansen-Goos}}\ and\ \bibinfo {author} {\bibfnamefont {J.~S.}\ \bibnamefont
  {Wettlaufer}},\ }\href {\doibase 10.1103/PhysRevE.81.031604} {\bibfield
  {journal} {\bibinfo  {journal} {Phys. Rev. E}\ }\textbf {\bibinfo {volume}
  {81}},\ \bibinfo {pages} {031604} (\bibinfo {year} {2010})}\BibitemShut
  {NoStop}%
\bibitem [{\citenamefont {Benatov}\ and\ \citenamefont
  {Wettlaufer}(2004)}]{Benatov2004}%
  \BibitemOpen
  \bibfield  {author} {\bibinfo {author} {\bibfnamefont {L.}~\bibnamefont
  {Benatov}}\ and\ \bibinfo {author} {\bibfnamefont {J.~S.}\ \bibnamefont
  {Wettlaufer}},\ }\href@noop {} {\bibfield  {journal} {\bibinfo  {journal}
  {Phys. Rev. E}\ }\textbf {\bibinfo {volume} {70}},\ \bibinfo {pages} {061606}
  (\bibinfo {year} {2004})}\BibitemShut {NoStop}%
\bibitem [{\citenamefont {Thomson}, \citenamefont {Benatov},\ and\
  \citenamefont {Wettlaufer}(2010)}]{Thomson2010a}%
  \BibitemOpen
  \bibfield  {author} {\bibinfo {author} {\bibfnamefont {E.~S.}\ \bibnamefont
  {Thomson}}, \bibinfo {author} {\bibfnamefont {L.}~\bibnamefont {Benatov}}, \
  and\ \bibinfo {author} {\bibfnamefont {J.~S.}\ \bibnamefont {Wettlaufer}},\
  }\href {\doibase 10.1103/PhysRevE.82.039907} {\bibfield  {journal} {\bibinfo
  {journal} {Phys. Rev. E}\ }\textbf {\bibinfo {volume} {82}},\ \bibinfo
  {pages} {039907} (\bibinfo {year} {2010})}\BibitemShut {NoStop}%
\bibitem [{\citenamefont {Wettlaufer}(1999)}]{Wettlaufer:1999}%
  \BibitemOpen
  \bibfield  {author} {\bibinfo {author} {\bibfnamefont {J.~S.}\ \bibnamefont
  {Wettlaufer}},\ }\href@noop {} {\bibfield  {journal} {\bibinfo  {journal}
  {Phys. Rev. Lett.}\ }\textbf {\bibinfo {volume} {82}},\ \bibinfo {pages}
  {2516} (\bibinfo {year} {1999})}\BibitemShut {NoStop}%
\bibitem [{\citenamefont {Rempel}\ \emph {et~al.}(2001)\citenamefont {Rempel},
  \citenamefont {Waddington}, \citenamefont {Wettlaufer},\ and\ \citenamefont
  {Worster}}]{Rempel:2001}%
  \BibitemOpen
  \bibfield  {author} {\bibinfo {author} {\bibfnamefont {A.~W.}\ \bibnamefont
  {Rempel}}, \bibinfo {author} {\bibfnamefont {E.~D.}\ \bibnamefont
  {Waddington}}, \bibinfo {author} {\bibfnamefont {J.~S.}\ \bibnamefont
  {Wettlaufer}}, \ and\ \bibinfo {author} {\bibfnamefont {M.~G.}\ \bibnamefont
  {Worster}},\ }\href@noop {} {\bibfield  {journal} {\bibinfo  {journal}
  {Nature}\ }\textbf {\bibinfo {volume} {411}},\ \bibinfo {pages} {568}
  (\bibinfo {year} {2001})}\BibitemShut {NoStop}%
\end{thebibliography}
\end{document}